\documentclass[aps,floats,superscriptaddress,twocolumn,pre]{revtex4}
\usepackage{amsmath,amssymb}
\usepackage{graphicx}
\usepackage{psfrag}
\usepackage{graphicx}

\def\beq{\begin{equation}}
\def\eeq{\end{equation}}
\def\bea{\begin{eqnarray}}
\def\eea{\end{eqnarray}}

\begin{document}

\title{Phase transitions and membrane stiffness in a class of asymmetric heterogeneous fluid membranes}

\author{Niladri Sarkar}\email{niladri.sarkar@saha.ac.in}
\author{ Abhik Basu}\email{abhik.basu@saha.ac.in}

\affiliation{Condensed Matter Physics Division, Saha Institute of
Nuclear Physics, Calcutta 700064, India}

\begin{abstract}{We propose a minimal model for miscibility phase
transitions (MPTs) in a class of asymmetric two-component
heterogeneous fluid membranes at equilibrium that generically
display both first and second order MPTs, controlled by the
interplay of asymmetry and heterogeneity. In the vicinity of the MPTs, the
membrane fluctuations are generally enhanced. However, the degree of
enhancement is found to depend sensitively on the asymmetry-heterogeneity
coupling. We argue that
experimental measurements of the membrane fluctuations at the MPTs should
provide physical information about the forms of the asymmetry-heterogeneity
couplings.}
\end{abstract}


\date{\today}

\maketitle

{\em Introduction:-} Miscibility phase transitions (MPTs) in
heterogeneous membranes at equilibrium are a subject of intense
research. Symmetric model lipid bilayers (i.e., lipid bilayers with
inversion symmetry) undergo an MPT from a high temperature ($T$)
homogeneous phase to a low temperature coexistence phase of liquid
disordered  and liquid ordered  domains~\cite{reviews}.
 Extensive experimental results
suggest that the  MPTs in symmetric model membranes are generically
second order in nature, belonging to the two-dimensional (2D) Ising
universality class~\cite{exp}. In contrast to their symmetric
counterparts, model heterogeneous asymmetric membranes are
distinguished by the lack of inversion symmetry and their
composition dependent {\em local spontaneous curvatures} $C_0$, that
affect the coarsening dynamics and equilibrium
shapes~\cite{asymexp}. Nonetheless, a general understanding of how
asymmetry affects the nature of the associated MPTs is still
lacking.

In this Letter we theoretically describe how the interplay
of asymmetry and inhomogeneity in a fluid membrane  affects the
 phases and the associated MPTs. To this
end, we construct a generic minimal model in the spirit of coarse-grained
Ginzburg-Landau approaches in terms of the local composition inhomogeneity and
curvature as the relevant thermodynamic variables. In order to focus on the
essential physical aspects of the problem, we consider a
tension-less single fluid membrane with { two-component}
heterogeneities. For the sake of simplicity and generality, we do
not distinguish between bilayer membranes and monolayer amphiphilic
films~\cite{andelman,andel2}. { Apart from its phenomenological
significance, our model is a good candidate to theoretically study 2D critical behaviour
on a fluctuating membrane.}


Our model displays a complex phase diagram with a rich variety of
phases and MPTs, including both second order transitions through
critical (CP)
 and tricritical  (TP) points, and first order transitions. Coupling constants that
 parametrise $C_0$ in our model appear as
control parameters. Associated membrane conformation fluctuations are
generically enhanced; however, the degree of enhancement can be controlled by
these tuning parameters. Furthermore, the magnitude and sign
 of $C_0$  in the ordered, phase-separated state in our model can be
tuned  at a given $T$ by controlling the parameters. Our
results
 demonstrate the importance of measuring the membrane fluctuations
 in experimental characterization of MPTs in asymmetric membranes.
 Our model is designed to capture the essential
physical consequences of nonlinear asymmetry-inhomogeneity
interactions and has few biological or microscopic details. Nonetheless,
considering the generality of our model, we expect the basic
features of our results, e.g., the significance of nonlinear
asymmetry-inhomogeneity couplings on the MPT and the nature of the
associated membrane fluctuations should be relevant  for non-linear curvature-composition interactions in
generic experiments on heterogeneous membranes and nonlinear aspects
of {\em lipid sorting} near phase transitions  in bilayer lipid
membranes~\cite{baum-nonlin}.

\begin{figure}[htb]
\includegraphics[height=1.4cm,width=8cm]{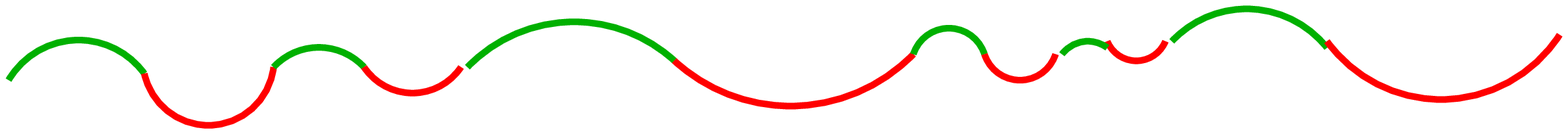}
\caption{(Color online) Schematic diagram of our model heterogeneous
amphiphilic membrane; red (curved upward)/green (curved downward) segments
represent A-excess/B-excess regions
with opposite $C_0$, respectively.} \label{model1}
\end{figure}

{\em Construction of our model:-} We describe inhomogeneity by a
single composition field $\phi ({\bf x})$. Physical interpretations
of $\phi$  depend on specific systems. For instance, for an
amphiphilic monolayers separating two distinct solvents (e.g., oil and water),
$\phi$ is the local difference between the
concentrations of the two types of lipids A and B; where as for
lamellae or vesicles made of bilayer membranes, it is the local
composition difference between the two layers of the
bilayer~\cite{andelman,andel2}, or, for a diffusing chemical in a
membrane, e.g., in {\em echinocytosis} of red blood cells, it is the
local density fluctuations of the diffusing
molecules~\cite{andel2,leibler,sheetz}. Naturally, the underlying
microscopic mechanisms behind the inversion asymmetry differ from
one system to another, e.g., different intrinsic spontaneous
curvatures of the two components in a two-component monolayer, the
composition difference between the inner and the outer
layers~\cite{andel2}, or, the preference of the intercalated
molecules for the tilted configurations of their surrounding
phospholipids ~\cite{leibler} in a bilayer. We adopt the standard
Ginzburg-Landau free energy functional for binary
mixtures~\cite{andelman,safran,chaikin}, useful near a critical point. 
 We consider $C_0(\phi)$ to be a generic nonlinear
function of $\phi$~\cite{lin1}, and for purposes of illustration
assume $C_0(\phi)=-C_0(-\phi)$, so that A and B-excess regions have
opposite local curvatures (assuming an amphiphilic membrane); see
Fig.~\ref{model1}.
 Next, assume the standard Helfrich-Canham
curvature energy for a
tensionless~\cite{surface-tension}, zero
thickness~\cite{surface1} membrane, parametrised by the bending modulus $\kappa$
and a single-valued height field $h({\bf
x},t)$, of the form $\kappa [\nabla^2 h - C_0 (\phi)]^2$ in the
Monge gauge~\cite{weinberg-book},  for planar fluid membrane
configurations. Evidently, the free energy functional $\mathcal F$
is invariant under  $h\rightarrow -h$ together with $\phi\rightarrow -\phi$. We have then,
  \bea \mathcal{F}&=&\int d^dx [{r \over 2}\phi^2 +
{1 \over 2}(\nabla\phi)^2+{u \over 4!}\phi^4 +{v \over
6!}\phi^6 \nonumber \\ &+&{\kappa \over 2}(\nabla^2h)^2
+\lambda\phi\nabla^2h +\lambda_1\phi^3\nabla^2h], \label{free}
 \eea
truncating to the lowest order nonlinear terms. Here $r\sim
(T-T_c)$, $T_c$ being the mean-field (MF) critical temperature
without any coupling to $h$. Coupling constants $u>0$ and $v>0$
 determine the strength of the local lipid-lipid interactions. The
$v$-term in $\mathcal{F}$ is required for thermodynamic stability as
we discuss below. The absence of invariance under the inversion of
$h$ for a given $\phi$ enforces asymmetry in the
model~\cite{helfrich1,leibler}.
 We have $C_0(\phi)=-(\lambda\phi + \lambda_1\phi^3)/\kappa$. Notice
 that the individual signs of $\lambda$ and $\lambda_1$ are
arbitrary: Signs of both $\lambda,\lambda_1$ may be flipped by
redefining $h\rightarrow -h$, or, $\phi\rightarrow -\phi$. The sign
of the product $\lambda\lambda_1$, however, cannot be transformed
away. In fact, the signature of $\lambda\lambda_1$ pertains to
important microscopic material properties: $\lambda\lambda_1
>0$ implies that the contributions of the microscopic
interactions between curvature and three or one lipid molecules to
$\mathcal F$ are mutually cooperative; on the other hand for
$\lambda\lambda_1<0$, these microscopic interactions are mutually
competing ~\cite{tricrit}. In the latter case, for $\lambda\sim
-\lambda_1\phi^2$ the two contributions roughly cancel out. This has
strong ramifications on the MPTs, as we find below. On dimensional
ground in 2D, following the arguments in Ref.~\cite{andelman},
$\lambda n_0 \sim \kappa H_0,\,\lambda_1 n_0^3 \sim \kappa H_0$,
where $n_0$ is the mean concentration and $H_0$ is a typical
microscopic (local) curvature, yielding $\lambda\sim \kappa
H_0\xi_0^2,\,\lambda_1\sim \kappa H_0 \xi_0^6$~\cite{curv}, with
$\xi_0$ being the microscopic correlation length scale.
Taking $\xi_0\sim 10^{-9}m$~\cite{andelman}, $\kappa\sim 100 k_BT$
($k_B$ is the Boltzmann constant) and $H_0\sim
10^6m^{-1}$, we find $\lambda\sim 10^{-10}k_BT m,\lambda_1\sim 10^{-46} k_BT m^5$.

{\em Mean-field analysis:-} We now construct a mean-field theory
(MFT)~\cite{chaikin} by minimizing $\mathcal F$ with respect to
(constant in MFT)  twice the negative of the mean curvature
$C=\nabla^2h$ and order parameter $m=\phi$. We find
\begin{eqnarray}
&&rm+\lambda C + 3\lambda_1 m^2 C + \frac{u}{3!}m^3 +\frac{v}{5!}m^5
=0,\label{mfm}\\
&&\kappa C +  \lambda m +\lambda_1 m^3 =0.\label{mfc}
\end{eqnarray}
For symmetric membranes $\lambda=0=\lambda_1$. Hence, $C=0$ for all
$T$ and $m=0$ for all $T\geq T_c$ and $\phi=m\neq 0$ for all $T<T_c$
with a second order transition at $T_c$, belonging to the Ising MF
universality class. Our results for an asymmetric membrane are far
richer in behavior: Eq.~(\ref{mfc}) shows that for an asymmetric
membrane in the ordered phase with a non-zero $m$ (a) $C\neq 0$, and
(b) the signatures of $C$, being dependent on that of $m$ not
surprisingly, are opposite in the A or B rich domains in the ordered
phase. Now eliminate $C$ in Eq.~(\ref{mfm}) by using Eq.~(\ref{mfc})
to
construct an {\em effective} Landau free energy ${\mathcal F}_e$:
\bea \mathcal{F}_e = {\tilde r \over 2}m^2 + {\tilde u}m^4 + {\tilde
v} m^6 , \label{efffree} \eea where $\tilde r=r-{\lambda^2 \over
\kappa}$, $\tilde u={u \over 4!}-{\lambda\lambda_1 \over \kappa}$
and $\tilde v={v \over 6!}-{\lambda_1^2 \over 2\kappa}$. Parameter
$\tilde r$ defines effective MF critical temperature $\tilde
T_c=T_c+{\lambda^2 \over \kappa}$. { Effective coupling constants}
$\tilde u$ and $\tilde v$ can be either positive, negative or zero,
separately or together. We assume $\tilde v>0$ always. Then,
$\mathcal F_e$ is identical to that for the normal superfluid
transition in liquid helium mixtures~\cite{graf}.

 Free energy (\ref{efffree}) allows for both first and second order transitions in the system, depending
upon the relative magnitudes and signatures of $\tilde r,\tilde
u$~\cite{chaikin}: (i)
 By construction when $\tilde r=0,\,\tilde u>0,\,\tilde v>0$,
 $\mathcal F_e$ admits a second order phase transition
for $m$ belonging to the MF Ising universality class. This holds for
all $\lambda,\lambda_1$, such that $\lambda\lambda_1 <0$, and also
for $\lambda\lambda_1 >0$ as long as $\tilde u >0$, (ii) For
sufficiently large $\lambda\lambda_1>0$, $\tilde u <0$ and the
$\tilde v$-term is then necessary for thermodynamic stability. For
this, the system undergoes a first order phase transition with
 $m=\pm \sqrt{{|\tilde u| \over 2\tilde v}}\neq 0$  at  $\tilde r^*=4\tilde u
m^2-6\tilde v m^4={\tilde u^2 \over 2\tilde v}>0$, yielding a
transition temperature $\tilde T_c^*=\tilde T_c+ {\tilde u^2 \over
2\tilde v}$. This meets the second order transition at $\tilde
r=0,\,\tilde u=0$, which defines a TP. The corresponding MF critical
scaling exponents belong to the MF  TP universality
class~\cite{chaikin}.


For an asymmetric membrane, $C=0$  in the disordered phase and in
general $C=C(m)=-(\lambda m +\lambda_1 m^3)/\kappa\neq 0$ in the
ordered phase. Hence,  with equal and opposite $C$ in the A and
B-rich domains, $C(m)$ changes sign as $m$ changes sign in the
ordered phase. Variation of $C(m)$ across the transition
temperature, that originates in the corresponding $T$ dependence of
$m$, may be used to delineate the nature of the transition: Below CP
or TP, it grows continuously from zero as $T$ decreases; in
contrast, it displays a jump, controlled by the jump in $m$, across
a first order transition~\cite{comment}. In addition, the magnitude and sign of $C$ may be
tuned in the ordered phase by controlling $\lambda$ and
$\lambda_1:$ $C=0$ along the {\em zero curvature line} $\lambda
+\lambda_1 m^2=0$ in the ordered phase. This requires
$\lambda\lambda_1 <0$ and hence $\tilde u>0$.


{\em Effect of small fluctuations:-} Consider now (small)
fluctuations in the disordered phase ($r>0$). Retaining terms up to
the harmonic order in $\mathcal F$ and integrating $\phi$ in the
partition function 
 ${\mathcal Z}=\int
{\mathcal D}h{\mathcal D}\phi \exp [-{\mathcal F}]$ ($k_BT=1$)
yields a wavevector ${\bf q}$-dependent effective bending modulus
\begin{equation}
\kappa_e=\kappa - \lambda^2/(r+q^2)\approx \kappa -
\frac{\lambda^2}{r}=\kappa_0<\kappa \label{kappaharm}
\end{equation}
 for $q\rightarrow 0$ at $r>0$. Thus, $k_e(q=0)=0$ at $\kappa=\lambda^2/r$ or $\tilde
 r=0$.
 Similarly in the
ordered phase define $\phi = m +\psi,\, \nabla^2 h=C +\delta c$,
where $m$ and $C$  are the solutions of Eqs.~(\ref{mfm}) and
(\ref{mfc}) in the ordered state and $\langle \psi\rangle = 0 =
\langle \delta c\rangle$. Expand $\mathcal F$ to the bilinear order
in $\psi$ and $\delta c$.  The terms linear in $\psi$ and $\delta c$
vanish, since $m$ and $C$ minimise $\mathcal F$ in the ordered
phase. We obtain
\begin{eqnarray}
{\mathcal F}_o&=&\int d^dx[\frac{r}{2}\psi^2
+\frac{1}{2}({\boldsymbol\nabla}\psi)^2 + \frac{6u}{4!}m^2\psi^2 +
15 \frac{v}{6!} m^4 \psi^2 \nonumber \\ &+& \frac{\kappa}{2} (\delta
c)^2 + \lambda \psi\delta c + 3\lambda_1 m^2 \psi \delta
c].\label{freeord}
\end{eqnarray}
Then, proceeding as before and integrating out $\psi$, we obtain an
effective free energy functional for the ordered phase that depends
on $\delta c$ only, and thence, assuming  $m^2=-3!(r/u)$  to the
lowest order in $\lambda,\lambda_1$ in the ordered phase, $\kappa_e$
in the ordered phase  given by
\begin{eqnarray}
\kappa_e(q)=\kappa-\frac{(\lambda + 3\lambda_1
m^2)^2}{-2r+q^2}\leq\kappa.\label{xyz}
\end{eqnarray}
In particular, along the zero spontaneous curvature line given by
$\lambda+3\lambda_1m^2=0$, $\kappa_e=\kappa$. Else, $\kappa_e <\kappa$.

{\em Role of the anharmonic fluctuations:-} Anharmonic fluctuation
contributions to $\kappa_e$, neglected above, should  dominate near
CP, due to the large (formally diverging in the thermodynamic limit)
critical fluctuations. As above, we need to integrate over $\phi$
(now retaining the anharmonic terms in $\mathcal F$) in $\mathcal Z$
near CP and obtain an effective free energy functional for $h$, and
thence extract $\kappa_e$ from there. Due to the
 anharmonicity, an exact integration is ruled out; instead perturbative calculations
 based on renormalisation group
(RG) calculations should be employed to systematically handle the large
(formally diverging in the thermodynamic limit (TL)) critical point
fluctuations~\cite{rgbook,chaikin}. This is a technically challenging
task. It is, however, easier to calculate the
height fluctuation correlator $\langle |h_{\bf q}|^2\rangle_\phi$ for a given
configuration of $\phi$. This is given by
\begin{equation}
 \langle |h_{\bf q}|^2\rangle_\phi = \int {\mathcal D}h \exp[{\mathcal
F}]\frac{1}{Z_\phi},\label{phiconfig}
\end{equation}
where $Z_\phi=\int {\mathcal D}h\exp[-{\mathcal F}]$. Equation
(\ref{phiconfig}) yields
\begin{equation}
 \langle |h_{\bf q}|^2\rangle_\phi = \frac{1}{\kappa q^4} +
|(\lambda\phi+\lambda_1\phi^3)_{\bf q}|^2\frac{1}{\kappa^2
q^4}>\frac{1}{\kappa q^4}.
\end{equation}
Thus, $\langle |h_{\bf q}|^2\rangle_\phi$ is always enhanced in the presence of
a given arbitrary configuration of $\phi$. Hence, the height
fluctuation correlator $\langle h_{\bf q}|^2\rangle$, averaged over all
possible configurations of $\phi$ with respect to the Boltzmann distribution
determined by ${\mathcal F}_0$ should be enhanced by $\phi$-fluctuations. Thus,
$\kappa_e <\kappa$ necessarily.

\begin{figure}[htb]
\centering
\includegraphics[width=8cm,height=6.2cm]{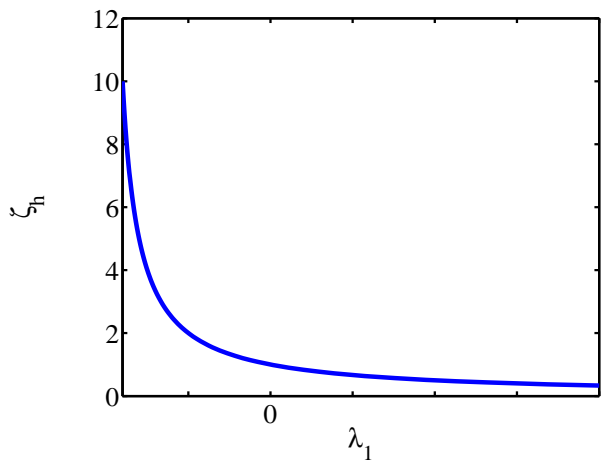} 
\caption{Schematic variation of $\zeta_h$ with $\lambda_1$ for a fixed $\lambda 
>0$. Clearly $\zeta_h$ is large for $\lambda\lambda_1 <0$ and gets smaller as 
$\lambda\lambda_1$ increases and becomes positive.}\label{pers}
\end{figure}

Actual enumeration of $\kappa_e$ can only be done perturbatively. This is
conveniently done by na\"ively expanding $\mathcal Z$ in powers of (assumed
small) $\lambda,\lambda_1,u$. This immediately yields,
\begin{equation}
 \kappa_e (q)=\kappa-\langle |\lambda\phi+\lambda_1\phi^3|_{\bf q}^2\rangle
\label{newres}
 \end{equation}
to $O(u^0)$. The last term in the rhs of (\ref{newres}) involves calculation of
higher order correlation functions of $\phi$, a difficult task by itself. In
order to get a quantitative sense of the nature of the correction to $\kappa$
in
(\ref{newres}), we resort to the Hartree approximation~\cite{chaikin} and
replace $\phi^3$ in
(\ref{newres}) by $3\langle\phi^2\rangle\phi$ (this amounts to ignoring
higher-order connected correlators of $\phi$). This yields
\begin{equation}
 \kappa_e(q)=\kappa- (\lambda+\lambda_1\langle\phi^2\rangle)^2\langle
|\phi_{\bf q}|^2\rangle,\label{newcorr}
\end{equation}
where $\langle\phi^2\rangle = \int \frac{d^dq}{(2\pi)^d}\langle |\phi_{\bf
q}|^2\rangle$ in $d$-dimensions that should be obtained self-consistently. 
In (\ref{newcorr}), contributions to $\kappa_e$ from connected higher order
correlations of $\phi$ are ignored. 
Ignoring self-consistency and noting that under spatial rescaling ${\bf
x}'=b{\bf x}$, coupling constants $\lambda$ and $\lambda_1$ and field $\phi$
scale as $b\lambda$, $b^{3-d}\lambda_1$ and $b^{(2-d)/2}\phi$,   all the
contributions to the corrections for $\kappa$ in (\ref{newcorr}) are equally
relevant at $2d$ in a scaling sense. For a system of linear size $L\sim 1/q_0$,
\bea
 \Delta\kappa_e(q_0)&=&\kappa_e(q_0)-\kappa\sim
-\frac{(\lambda +\lambda_1\langle\phi^2\rangle)^2}{\tilde r+q_0^2}\rightarrow
-\frac{(\lambda +\lambda_1\langle\phi^2\rangle)^2}{q_0^2}\nonumber \\
&&\sim - (\lambda +\lambda_1\langle\phi^2\rangle)^2L^2, \eea
near the critical point of MPT. Evidently, for a large enough $L$, 
$\kappa_e(L)=0$. This allows us
to define a {\em persistence length} $\zeta_h$ for the membrane
conformation fluctuations by $\kappa_e(\zeta_h)=0$, giving
\begin{equation}
\zeta_h\sim\sqrt{\frac{\kappa}{[\lambda+\lambda_1\langle\phi^2\rangle]^2}
}.\label{zetah}
\end{equation}
Thus, if $\lambda$ and $\lambda_1$ are of the same sign, then
$\Delta\kappa_e$ is increased in magnitude and $\zeta_h$ is decreased. In
contrast, when they are of opposite signs, $\Delta\kappa_e$ is reduced and
$\zeta_h$ is enhanced; see Fig.~\ref{pers} for schematic variation of 
$\zeta_h$ with $\lambda_1$ for a fixed $\lambda>0$. 
Accordingly, the membrane fluctuates more for
$\lambda\lambda_1>0$ (``more soft'') than for $\lambda\lambda_1 <0$ (``less
soft''). Since $\lambda$ and $\lambda_1$
are free parameters in our model, it is possible to make $\Delta k_e$
vanishingly small by choosing $\lambda\sim -\lambda_1\langle\phi^2\rangle$.
Within our analysis, this makes $\zeta_h$ diverging~\cite{geomnew}. The
corresponding  anharmonic fluctuation corrections
to $\kappa_e(q)$ in the ordered phase  close to CP to the lowest order in
$\lambda,\lambda_1$
\begin{equation}
\kappa_e (q) \approx \kappa - \frac{(\lambda+3\lambda_1
m^2+3\lambda_1\langle\psi^2\rangle)^2}{-2r+q^2},\label{kappeord}
\end{equation}
where $\langle\psi^2\rangle=\int \frac{d^2q}{(2\pi)^2}\langle
|\psi({\bf q})|^2\rangle$. Thus, $\Delta\kappa_e <0$ in general; its magnitude
may be controlled (can be increased or decreased) by tuning
the magnitudes and signs of $\lambda,\lambda_1$ as for above CP. In
contrast, across a first order transition, fluctuation effects should be
negligible. Hence, by using (\ref{xyz})
\begin{equation}
\Delta\kappa_e(q=q_0)=-\frac{(\lambda + 3\lambda_1 m^2)^2}{\tilde
r^*+q_0^2},\label{firstkappa}
\end{equation}
 $m^2=0,|\tilde u|/(2\tilde v)$ in the
disordered and ordered phase respectively. Hence, there is a jump in
$\kappa_e$ across a first order transition.
For a first order
transition, $\lambda\lambda_1
>0$, and hence, both $\lambda$ and $\lambda_1$ are either positive
or negative. Thus, $\Delta\kappa_e (q=q_0) <0$ necessarily for a
first order transition. Further, (\ref{firstkappa}) reveals a jump
in $\kappa_e$, due to the jump in $m$. Overall, thus we see that the
harmonic composition fluctuation contributions to $\kappa_e$ always
reduces $\kappa_e$. The corresponding anharmonic contribution may
further reduce $\kappa_e$, or, may partly suppress reduction of
$\kappa_e$ by the composition fluctuations at the harmonic order.
This can be argued heuristically from the form of (\ref{free}). When
$\lambda_1=0$, Gaussian fluctuations of $\phi$ is known to soften
the membrane as evident from (\ref{kappaharm}), due to a possible
reduction in free energy in a given bent configuration by adjusting
$\phi$-fluctuations. Noting that for $\lambda\lambda_1 <0$,  free
energy reduction due to the fluctuations of $\phi$ at the Gaussian
order is compensated by  a free energy cost due to the
$\lambda_1$-term. These two contributions (\ref{free}) should
balance when $\lambda\sim -\lambda_1\phi^2$. This explains the
mutually cooperative (competing) nature of the contributions to
$\kappa_e$ from the composition fluctuations at the harmonic and
anhamornic orders for $\lambda\lambda_1 > (<) 0$.

\begin{figure}[htb]
\centering
\includegraphics[width=8cm,height=6.2cm]{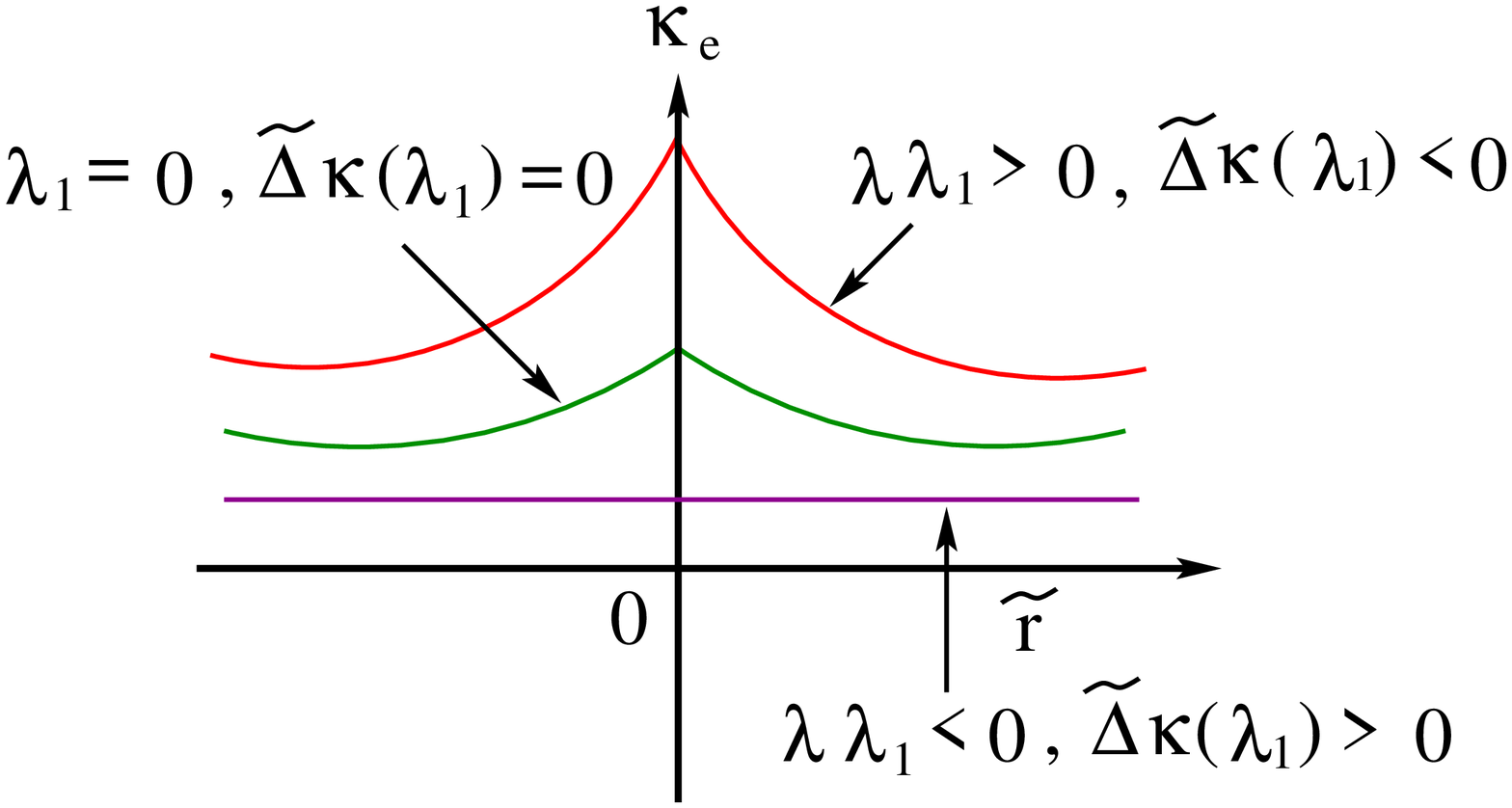}\\\includegraphics[
width=8cm
,height=5cm]{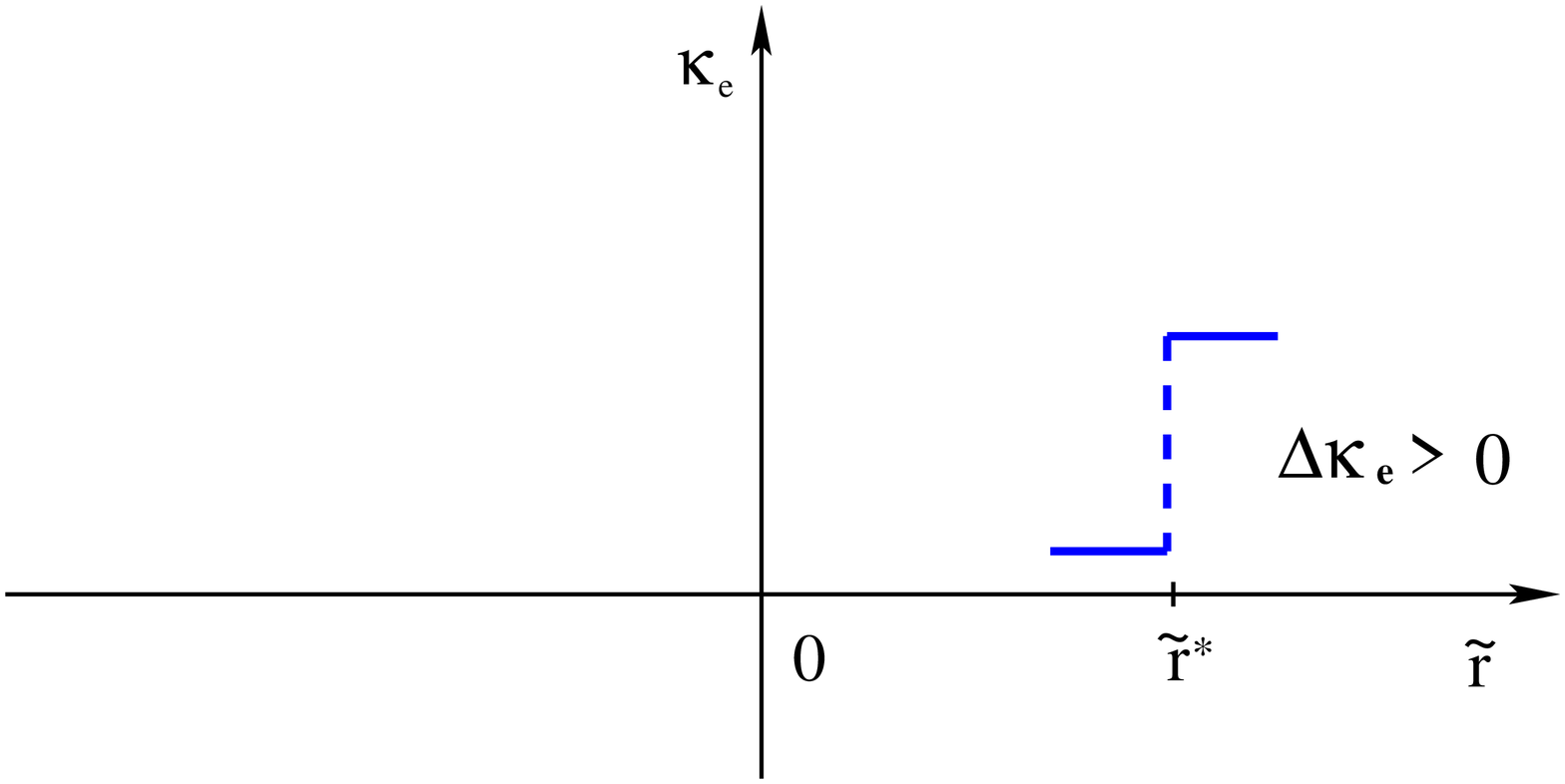} \caption{(Color online) Schematic
variations of $\kappa_e$ with $\tilde r$:
(top) Continuous across second order MPT with
$\lambda\lambda_1>0$ ($\tilde\Delta\kappa <0$, "membrane more soft", red line),
$\lambda_=0$ ($\tilde\Delta\kappa=0$, green line, nonlinear 
curvature-heterogeneity interaction vanishes) and
$\lambda\lambda_1
<0$ ($\tilde\Delta\kappa >0$, "membrane less soft", magenta line); (bottom) 
discontinuous 
across first
order transition at $\tilde r=\tilde r^*$; broken line indicates
$\Delta\kappa_e$ with no significant $L$-dependence (see text).}
\label{phasediag2}
\end{figure}

Considering that $\kappa$ has no significant $T$-dependences, measurements of
$\kappa_e$ as a function of $T$ yield information about $\lambda$ and
$\lambda_1$. For instance, for a second order MPT, measuring $\kappa_e$
versus $T$ above CP and using (\ref{kappaharm}) yields $\lambda$. Similar
procedure below CP, now with the knowledge of $m^2$ (use MFT results or measure
experimentally) yields $\lambda_1$. Similarly, across a first order
transition measuring
$\Delta\kappa_e$ as a function of $T>\tilde T_c^*$ yields
$\lambda^2$; see Eq.~(\ref{firstkappa}). Measuring the same quantity for
$T<\tilde T_c^*$ with
the knowledge of $\lambda$ obtained as above then yields
$\lambda_1m^2$. Use then $m^2=|\tilde u|/(2\tilde v)$ (or measure it
separately) below the
MPT to obtain $\lambda_1$.
For a first order
transition, $\lambda\lambda_1
>0$, and hence, both $\lambda$ and $\lambda_1$ are either positive
or negative. We classify the
membrane fluctuations near the associated MPTs according to the nature of the
corresponding MPTs - first order ($\lambda\lambda_1 >0,\,\tilde u <0$), second
order with cooperative asymmetry-heterogeneity interactions ($\lambda\lambda_1
>0,\,\tilde u>0$, ``membrane more soft'') and second order with 
competing asymmetry-heterogeneity
interactions ($\lambda\lambda_1
<0,\,\tilde u>0$, ``membrane less soft''). In the second and third cases, 
$\tilde\Delta\kappa = \kappa_e(\lambda_1)-\kappa_e(\lambda_1=0)$ is negative 
and positive, respectively. Thus, measurements of $\tilde\Delta\kappa$ yield 
information about the sign of $\lambda\lambda_1$. Notice that at the level of 
our analysis based on the Hartree
approximation, the last two cases differ only in the quantitative degree of
membrane fluctuation enhancement.
Variations of $\kappa_e$ with $\tilde r$ are
schematically shown in Fig.~(\ref{phasediag2}) across second order  and first
order transitions.


{\em Phase diagrams:-} Our results are summarised in schematic
phase diagrams (\ref{phasediag}-\ref{phasediag1}) below.  In
Fig.~(\ref{phasediag}) different types of transitions displayed by
our model (within our MF analysis) and the corresponding changes in
$\kappa_e$ across the transitions are marked in the $\tilde
r-\lambda$ plane ($\lambda_1>0$). The locations of TP given by the
condition $\tilde u=0$, and the line of zero spontaneous curvature,
given by $C(m)=0$ for particular choices of $\lambda,\lambda_1$ (in
the ordered phase), are shown. Figure~(\ref{phasediag1}) shows what
type of MPTs are expected for a given set of $(\lambda,\lambda_1)$
as temperature is lowered. The two (curved) lines of TP are given by
 $\tilde u=0$, or,
$\lambda\lambda_1/\kappa=u/4!$. For $\lambda\lambda_1
>u\kappa/4!$ (regions marked A), $\tilde u<0$, and hence corresponds to 
first order transitions. In regions marked B, on the
other hand, $\tilde u>0$ with $\lambda\lambda_1 >0$ and hence they
correspond to second order transitions with a very soft membrane
across the transition. In contrast, $\tilde u >0$ with
$\lambda\lambda_1 <0$ in regions marked C; these describe second
order transitions with ``less soft'' membranes across the
transitions.
\begin{figure}[htb]
\includegraphics[width=8cm,height=7.2cm]{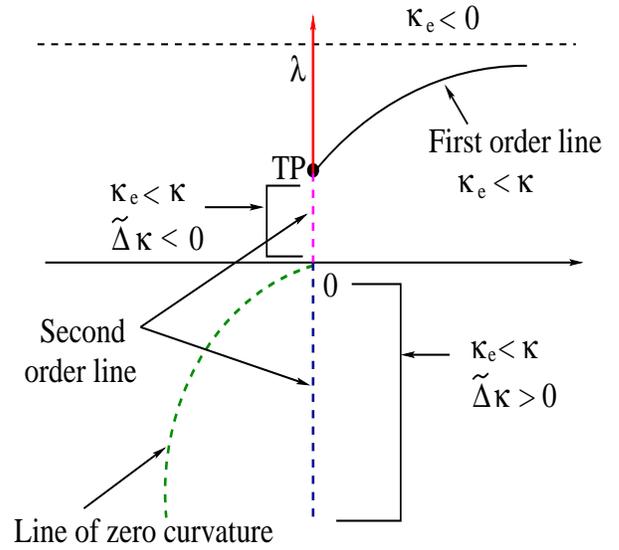}
\caption{(Color online) Schematic phase diagram in the $\tilde
r-\lambda$ plane ($\lambda_1 >0$). First and second order lines with
the change in $\kappa_e$ across the transitions and TP are marked. A
line of zero curvature is schematically drawn.}  \label{phasediag}
\end{figure}
\begin{figure}[htb]
\includegraphics[width=8cm,height=5.2cm]{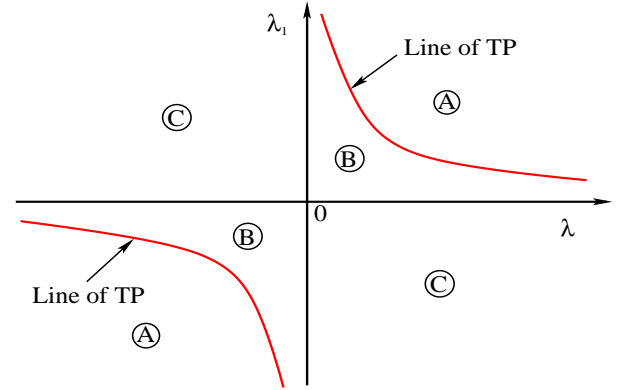}
\caption{(Color online) The nature of transitions in the different
regions of the $\lambda-\lambda_1$ plane:  Systems with
$\lambda,\lambda_1$ belonging to the regions marked A, B and C
undergo first, second ($\lambda\lambda_1>0,\tilde u>0,\tilde\Delta\kappa<0$) 
and second
($\lambda\lambda_1<0,\tilde u >0,\tilde\Delta\kappa >0$) order transitions 
respectively; red curved
lines are the locations of
TPs
($\tilde u=0$).}\label{phasediag1}
\end{figure}


{\em Summary and outlook:-} Thus, we show how the interplay between
composition and asymmetry determine the ensuing phases and phase
transitions, and generic suppression of membrane fluctuations in
asymmetric membranes, which are markedly more complex than
symmetric membranes. Furthermore, the degree of enhancement can be controlled
by the coupling constants that define the coupling between the local asymmetry
and heterogeneity.  
Since $\lambda,\lambda_1 $ have very different dependences on
$\xi_0$, performing experiments on model asymmetric heterogeneous
membranes with different sizes of the constituent lipid molecules
should be a promising route to test our results experimentally. For
instance, if $\xi_0$ changes by 10\%, $\lambda,\lambda_1$ change by
about 20\% and 60\%, respectively. Thus shows that by varying $\xi_0$, the
effective stiffness of an asymmetric membrane can vary significantly. Our model
demonstrates the
significance of the sign of the product $\lambda\lambda_1$.
Complementary to our analysis above, phases and phase transitions in
(\ref{free})  may also be tested in numerical simulations of
suitably constructed equivalent lattice-gas based 2D lattice
Hamiltonian with an Ising spin degree of freedom $S_{ij}$, a
discrete analogue of $\phi({\bf x})$ here ($(i,j)$ is the coordinate
of a point on the 2D lattice), and a discretised local curvature of
the lattice $h_{ij}$ (see, e.g., Ref.~\cite{wallace}). Discrete
analogues of the $\lambda$- and $\lambda_1$-terms may be constructed
by coupling $S_{ij}$ with $h_{ij}$ appropriately. Additionally,
atomistic models for asymmetric lipid bilayers that may correspond
to the coarse-grained free energy (\ref{free}) may be constructed by
introducing local many-body interactions involving three atoms (or
interaction sites) and the curvature and controlling its sign. In
turn, knowledge about $\tilde \lambda$ should yield useful
information about the underlying microscopic interactions. We look
forward to synthesis of model asymmetric lipid membranes which may
be used in experiments by standard, e.g., fluorescence, methods
(see, e.g., Ref.~\cite{exp}) to study the properties elucidated
above. Membrane fluctuations and their dependences on $\lambda,\lambda_1$
may be investigated by standard experimental methods, e.g., optical
interferometric
methods~\cite{flicker}.  The
number of tunable parameters in our model is similar in number in
analogous models for symmetric inhomogeneous membranes (see, e.g.,
Ref.~\cite{ayoton}. Nevertheless, the MPTs in our asymmetric
inhomogeneous model membranes are starkly different from those for
symmetric ones. This highlights the crucial role of asymmetry.

Our results are strongly related to the symmetry properties of
(\ref{free}), i.e., $C_0(\phi)=-C_0(-\phi)$, or, equivalently,
symmetry under the joint inversion $(h,\phi)\rightarrow (-h,-\phi)$.
If the molecular curvatures of lipid molecules A and B are
different, then this symmetry property will not hold good. Insisting
on modeling asymmetric membranes with no particular symmetry under
inversion of $\phi$, we can write down a generalised
$C(\phi)=\lambda\phi + \lambda_1 \phi^3 + \lambda_2\phi^2$
(truncating up to $\phi^3$). The new  $\lambda_2$-term evidently
breaks the symmetry under the inversion $(h,\phi)\rightarrow
(-h,-\phi)$. Model asymmetric membranes with a non-zero $\lambda_2$
but with $\lambda=0=\lambda_1$ have been studied theoretically in
Ref.~\cite{niladri1}, which illustrates the possibility of both
first and second order transitions and generic enhancement of
membrane fluctuations near the second order transitions. In the more
general case with non-zero $\lambda,\lambda_1$ and $\lambda_2$, we
expect a combination of the results from this work and
Ref.~\cite{niladri1} to emerge, including possibly a more complex
phase diagram. While a full analysis is beyond the scope of the present
work, nonetheless, the results of Ref.~\cite{niladri1} together with those here
strongly highlights the generic nature of enhancement of membrane fluctuations
and the possibilities of both first and second order MPTs.

Some technical comments are in order now. In writing $\mathcal F$
[Eq.~(\ref{free})] we have neglected the geometric nonlinearities
which arise from expanding $h({\bf x})$ about the perfectly flat
base plane in the Monge gauge, e.g., area element
$dS = d^2x \sqrt{1+({\boldsymbol\nabla}h)^2}\simeq d^2x [1+
({\boldsymbol\nabla}h)^2/2],$
and the  mean curvature
$c_{mean}=-\nabla^2 h +\frac{1}{2}\nabla^2h ({\boldsymbol
\nabla}h)^2 + \partial_i h\partial_j h \partial_{ij}h$
for small fluctuations in $h$; $i,j=x,y$. Inclusion of the above in
(\ref{free}), generate additional nonlinear terms. Again straight
forward scaling analysis near CP directly yields that these
geometric nonlinearities are all {\em irrelevant} (in a scaling
sense) in the presence of the couplings $u$ and $\lambda_1$. Hence,
to the leading order, the geometric nonlinearities should be
subleading to the existing nonlinearities in (\ref{free}). This
justifies omission of the geometric nonlinearities in our analysis
above.  Furthermore, there are additional symmetric (invariant under
$h\rightarrow -h$ and $\phi\rightarrow-\phi$) nonlinear terms
coupling $h$ and $\phi$, e.g., $\phi^2(\nabla^2 h)^2$ (ignored
here), which are of thermodynamic origin and generically present for
both symmetric and asymmetric heterogeneous membranes. This is,
however, irrelevant (in a scaling sense), similar to the geometric
nonlinearities. Thus, this term leaves the critical properties of
MPT and the associated membrane fluctuations in our model asymmetric
membrane unaffected. Within our Hartree approximation, there
are little qualitative differences between $\lambda\lambda_1>0$ and
$\lambda\lambda_1 <0$ (keeping $\tilde u>0$ for both), except for
the degree of enhancement of the membrane fluctuations near the
associated MPT. Whether this is indeed the case or whether there are indeed 
significant qualitative differences between the two cases that are missed by 
our low-order perturbation theory should be investigated by more elaborate 
calculations. Our
results suggest that in the special case with $\lambda\lambda_1 <0$
in (\ref{free}), when the harmonic and anharmonic composition
fluctuation contributions to $\kappa_e$ nearly mutually cancel to
the leading order, the geometric and other symmetric nonlinearities
mentioned above should become relevant, and the long wavelength
properties of an asymmetric heterogeneous membrane described by
(\ref{free}) should be identical to a symmetric heterogeneous
membrane, as described in Ref.~\cite{tirtha-membrane}. Whether or
not this actually happens can be determined by more detailed
calculations that are beyond the scope of this work. Nonetheless, we
can  conclude with reasonable confidence that near the MPTs, an
asymmetric heterogeneous membrane is likely to be generally softer
than the corresponding symmetric heterogeneous membrane with the
same bare bending modulus. In addition, in symmetric membranes
($\lambda\lambda_1=0$) the critical behaviour of $\phi$ will be
controlled by the $u$-term, and hence, Ising-like. This establishes
the 2D Ising universality for the MPTs in symmetric heterogeneous
membranes, consistent with the known experimental
results~\cite{exp}.

Studies on the possibility of budding transitions (see, e.g.,
Ref.~\cite{kumar}) in our model and its relation to the divergence
or suppression of membrane fluctuations at CPs should be
interesting. Time-dependent phenomena in our model, e.g., dynamic
scaling and growth of order after a temperature quench through CPs,
TPs or first order transitions should be of interest from
experimental point of views. Our work should be  useful in the
context of static and dynamic properties of phase transitions in
Langmuir monolayers of polar molecules~\cite{andel}.  We hope our
results will stimulate further theoretical and experimental works
along these directions.


{\em Acknowledgement:-} We thank the anonymous Referee for his/her comments 
and suggestions. One of the authors (AB) wishes to thank  the
Max-Planck-Society (Germany) and the Department of Science and
Technology/Indo-German Science and Technology Centre (India) for
partial financial support through the Partner Group programme
(2009).


\begin{thebibliography}{99}
\bibitem{lin1} Given our wisdom gathered from extensive studies on critical phenomena,
 symmetry-allowed nonlinear terms in $C_0$ may become relevant near
 CPs.
\bibitem{tricrit}  This is formally similar to
mutually competing or cooperative $\Psi^4$ and $\Psi^6$ terms in a
Ginzburg-Landau descriptions for magnetic spin models having a
tricritical point with $\Psi$ being the local spin~\cite{chaikin}.
\bibitem{comment} Monge gauge holds for small fluctuations in a planer membrane
with small $C_0(\phi)$. Since $m$ is small just below CP/TP, $C(m)$
should also be small, thus validating  Monge gauge. In a first order
transition, $C(m)$ is finite just the transition temperature. Our
formulation then holds for small $C(m)$, tunable by
$\lambda,\lambda_1$. We generally expect the membrane to remain
planar on average sufficiently close to the transition temperature,
for which our theory is valid. Our theory cannot predict the overall
shape for large $C(m)$ deep in the ordered phase.
\bibitem{geomnew} In this case, several symmetric nonlinearities of both
geometric and thermodynamic origin which are neglected here, should become
relevant. Their interplay will determine $\zeta_h$; see below for more
discussions; see also Ref.~\cite{}.
\end{thebibliography}
\end{document}